# Sequential tasks performed by catalytic pumps for colloidal crystallization

*Ali Afshar Farniya[1], Maria J. Esplandiu[1,2*], Adrian Bachtold[3]*

[1]ICN2 – Institut Catala de Nanociencia i Nanotecnologia, Campus UAB, 08193 Bellaterra (Barcelona), Spain

[2]CSIC – Consejo Superior de Investigaciones Científicas, ICN2 Building, Campus UAB, 08193 Bellaterra (Barcelona), Spain

[3] ICFO-Institut de Ciencies Fotoniques, Mediterranean Technology Park, 08860 Castelldefels (Barcelona), Spain

ABSTRACT

 Gold-platinum catalytic pumps immersed in a chemical fuel are used to manipulate silica colloids. The manipulation relies on the electric field and the fluid flow generated by the pump. Catalytic pumps perform various tasks, such as the repulsion of colloids, the attraction of colloids, and the guided crystallization of colloids. We demonstrate that catalytic pumps can execute these tasks sequentially over time. Switching from one task to the next is related to the local change of the proton concentration, which modifies the colloid zeta potential and consequently the electric force acting on the colloids.

*Corresponding Author: MariaJose.Esplandiu@cin2.es

1. INTRODUCTION

In the last years there has been a growing interest in the study of active matter. These systems, which work far from equilibrium driven by a constant input of energy, exhibit various interesting and surprising phenomena such as collective dynamics, complex self-organization, and the emergence of large-scale coherent structures [1-12]. Biology provides a wealth of archetypes of active systems that consume chemical energy to self-organize into complex structures and to perform different collective tasks. Some examples include cytoskeleton motors, self-propelled microorganisms, biofilm patterning, bacterial colonies [13-26] and even at larger scales swarming phenomena, bird flocks, and fish schooling [27-29]. The tasks achieved by biological active systems are much more sophisticated than those of active systems fabricated thus far by humans.

Artificial swimmers such as self-propelled active colloids and catalytic Janus particles also provide nice examples of collective motility, swarming and dynamic self-assembly [1,3,5,6,8-11,30-44]. Microfabricated catalytic pumps can be used to manipulate particles in solution and to form crystals of colloids [45-50]. These active swimmers and pumps are driven by catalytic reactions on their surface which triggers electro-hydrodynamic forces [5,8,30-52]. The growth of colloidal crystals with swimmers and pumps is an interesting alternative to the more traditional colloidal self-assembly triggered by external fields [53-65]. However, artificial active systems have accomplished tasks that are rather simple thus far. Usually, artificial active systems effectuate only a single task that does not change in time in an autonomous way.

In this paper, we use microfabricated catalytic pumps to manipulate silica colloids by means of fluid flow and electric field. Interestingly, we have found that catalytic pumps can perform a sequence of different tasks over time. First, pump disks repel colloids far away. Second, pump disks direct colloids towards them. Third, pump disks assemble colloids in a crystal around them. We correlate the change between the first and the second task to the variation of the proton concentration which greatly affects the zeta potential of the colloids. The formation of the colloidal crystal is assisted by the electro-hydrodynamic forces of the pump. We observe how the colloids rearrange themselves in the crystal in order to heal defects.

2. EXPERIMENTAL SECTION

Catalytic pumps consist of 20-50 μm diameter Pt disks patterned on gold films. Pumps are fabricated using standard electron-beam lithography techniques. Pumps are subjected to one minute of oxygen plasma cleaning (360 Watt) to remove residual PMMA/organic contamination and activate the surface as reported before [48]. An 8 mm diameter and 0.12 mm thick gasket-like spacer (Invitrogen) is placed on top of the wafer patterned with micropumps. A 1 wt% hydrogen peroxide solution containing negatively charged colloids is added to the vacant space created by the gasket. The wafer is immediately capped with a thin glass cover. The negative colloids are 1.5 μm diameter silica spheres (Kisker Biotech GmbH & Co) with a zeta potential ($\xi$) of -83.5 mV. The concentration of colloids is about $10^9$ particles/mL. The motion of particles is optically recorded with 5 frames per second rate and analyzed with the Diatrack software to determine their velocity. The zeta potential of the colloids and their variation in different pH solutions are carried out with a Zetasizer Nano-ZS (Malvern Instruments) based on electrophoretic light scattering.

## 3. RESULTS AND DISCUSSION

Hydrogen peroxide is used as the chemical input to trigger the catalytic actuation. The chemical fuel decomposes at both metal structures, one of them acting as anode and the one other one as cathode (Fig. 1a)

Reaction at the anode :  $H_2O_2 \rightarrow O_2 + 2H^+ + 2\ e^-$  (1)

Reaction at the cathode:  $H_2O_2 + 2H^+ + 2\ e^- \rightarrow H_2O$  (2)

Previous results demonstrated that oxygen plasma cleaning of the surface is mandatory to activate motion. Under these conditions platinum disk acts as the cathode and the gold surface as the anode in presence of $H_2O_2$ [48]. The electrochemical reaction generates an excess of protons at the anode which is consumed at the cathode. As a result, an electric field pointing towards the Pt disk is generated, which drives a fluid flow in the same direction (Fig. 1 a).

The motion of colloids can be basically ascribed to two contributions, one coming from the electrophoretic force ($v_{eof} = \varepsilon E_r \xi_c / \eta$) and the other arisen from the fluid flow ($v_f$). The fluid moves because of electro-osmosis, so that $v_f = -\varepsilon E_r \xi_{Au} / \eta$. The radial velocity of colloids is given by

$$v_r = v_{eof} + v_f = \varepsilon E_r (\xi_c - \xi_{Au})/\eta \qquad (1)$$

where $E_r$ is the radial electric field, $\varepsilon$ is the fluid permittivity, $\eta$ the fluid viscosity, and $\xi_c$ and $\xi_{Au}$ are the zeta potentials of colloids and the Au surface. Previous measurements allowed us to estimate the zeta potential of the Au film $\xi_{Au} = -33$ mV [48].

When $H_2O_2$ containing negatively charged silica colloids ($\xi_c$ = -83.5 mV) is placed onto catalytic micropumps, a region free of colloids forms around the Pt disk (Figs. 1b-e). Colloids remain more than 20 μm away from the disk edge. This region free of colloids is consistent with a strong radial electric force that repels negatively charged colloids from the pump disk.

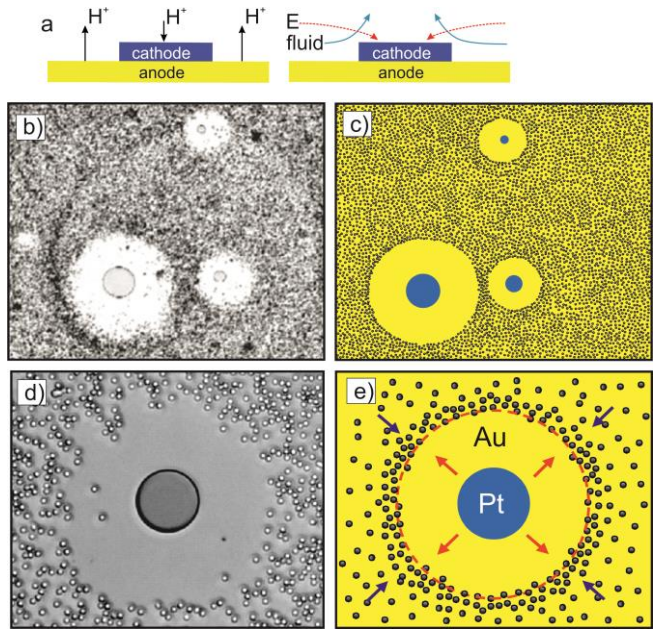

**Figure 1.** a) Schematics representing the production and consumption of protons, the electric field line and the fluid flow. (b-e) Optical microscopy images and schematics of pumps and colloids when $H_2O_2$ is put on the surface of the device. Colloids are repelled from the pump disk. In e) the red arrows indicate the electric force acting on the colloids, whereas the blue arrows indicate the fluid flow.

However this repulsive area is only formed at the very first stage of the experiment. Many other interesting features start emerging in time as it is outlined in Fig. 2. After the initial repulsion, silica colloids start aggregating into small clusters (Fig. 2, a-b). At the same time, these small clusters as well as individual colloids start moving towards the Pt disk (Fig. 2, c-d). Once arrived at the pump disk, colloids assemble themselves in a crystal (Fig. 2, e-f).

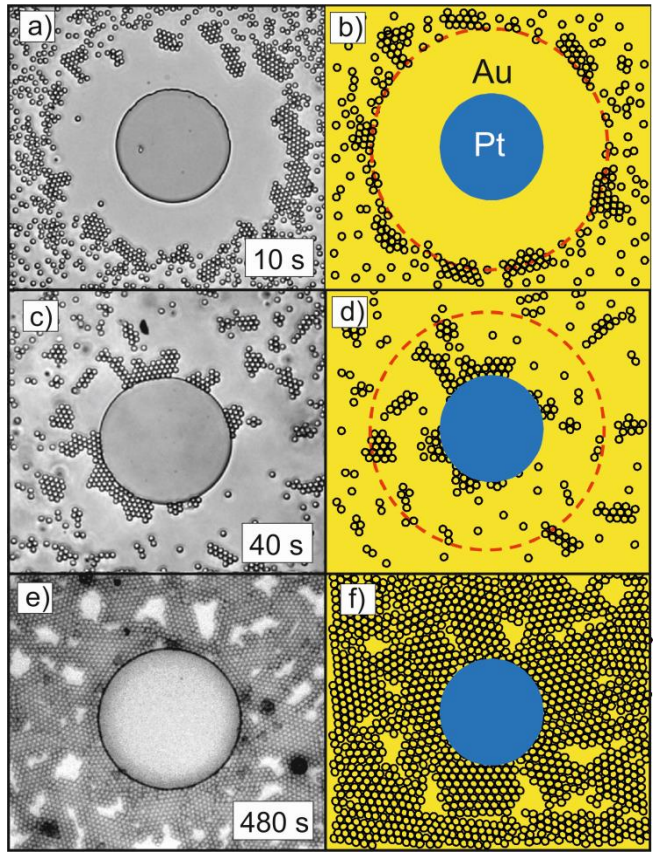

**Figure 2**. Optical microscopy images and schematics of pumps and colloids. (a-b) Cluster formation far away from the pump disk. (c-d) Motion of small clusters and individual colloids towards the pump disk. (e-f) Colloidal crystal formation around the pump disk

The formation of small clusters in Figs. 2a,b are attributed to the production of protons at the Au surface. To show this, we fabricate Pt-Au micropumps patterned near a region with a silicon oxide surface (Fig. 3a). Colloids do not assemble together in the region with the silicon oxide surface. They only feature Brownian motion. By contrast, the formation of clusters occurs over the gold surface, the region where protons are produced. This suggests that the cluster formation is correlated to the proton production. To gain more insights on this issue, we monitor the dispersion of colloids above a gold surface for different proton concentrations in absence of hydrogen peroxide (Figs 3b-d). Colloids do not assemble together at high pH, but they start to form small clusters at pH ≈ 5.5. This pH is close to the one estimated on the gold surface when the pump is actuated with $H_2O_2$ [48].

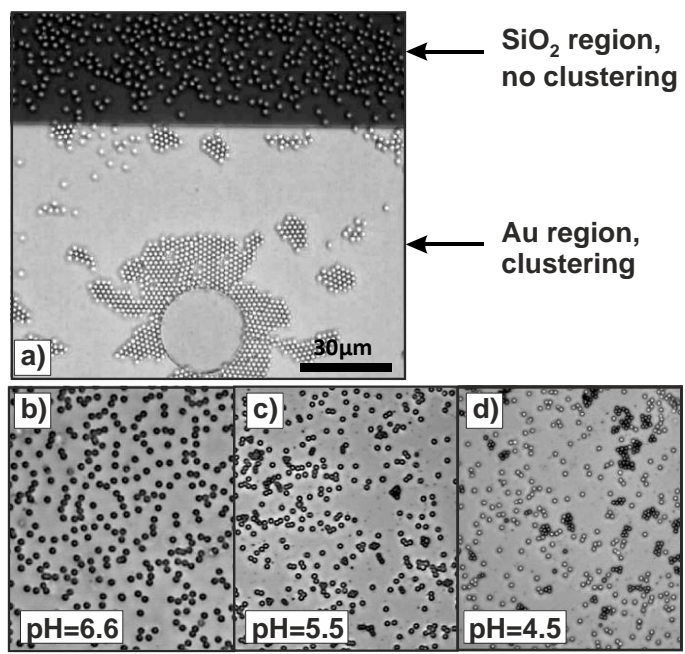

**Figure 3.** a) Optical microscopy image of a micropump patterned near a region with a silicon oxide surface. (b-d) Optical snapshots of the dispersion of colloids at different pHs in absence of hydrogen peroxide. The clustering occurs below a pH of around 5.5.

In Fig. 4a, we measure the variation of the zeta potential of silica colloids as a function of pH. The colloid zeta potential becomes less negative upon decreasing the pH. Overall, all these experiments indicate that the formation of small clusters is due to the change of the zeta potential of colloids subjected to the production of protons at the Au surface. Indeed, the continuous generation of protons at the surface might protonate the negative oxygen moieties of the silica, so that the charge of the colloid is reduced, and the interaction between the colloids is promoted.

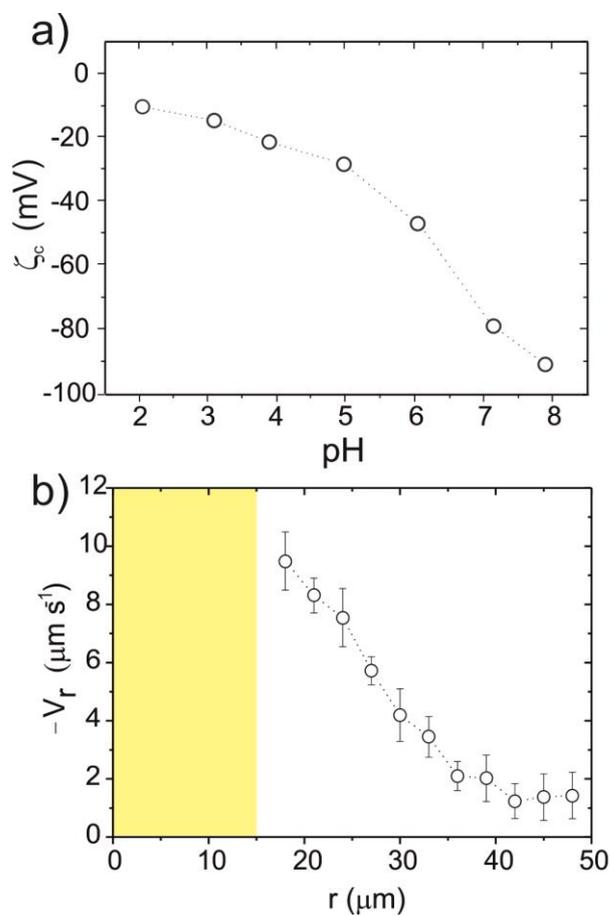

**Figure 4.** a) Variation of the zeta potential of colloids as a function of pH. The measurements are done by preparing different pH solutions in a 1 mM buffer phosphate. b) Averaged radial velocity of colloids as a function of the radial coordinate $r$.

The motion of the colloids towards the pump disk in Fig. 2c,d is attributed to the production of protons at the Au surface as well. Because of the change of the zeta potential induced by protons, the electric repulsion of the colloids from the pump disk is lowered, allowing colloids to move towards the disk. The radial velocity of colloids, obtained by averaging 10 different trajectories, increases as colloids approach the pump disk (Fig. 4b). This increase of the velocity at the pump disk is similar to what was reported for positively charged colloids and quasi-neutral colloids [48]. This further supports the change of the zeta potential of silica colloids.

More than fascinating is the real-time monitoring of the colloidal self-assembly at the disk edge. The colloids stabilize at the edge of the pump disk building a hexagonal crystal which expands over the gold surface. The incoming small clusters change form and reorient themselves to better fit in the lattice of the growing colloidal crystal (red arrows in Figs. 5 a,b). In addition, the defects of the crystal tend to disappear over time (arrows in Figs. 5 c,d). These effects are attributed to the force from the moving fluid and the thermal energy that helps colloids to find the more coordinated positions. We do not know at present the microscopic origin of why the colloidal crystal starts to grow from the edge of the Pt disk.

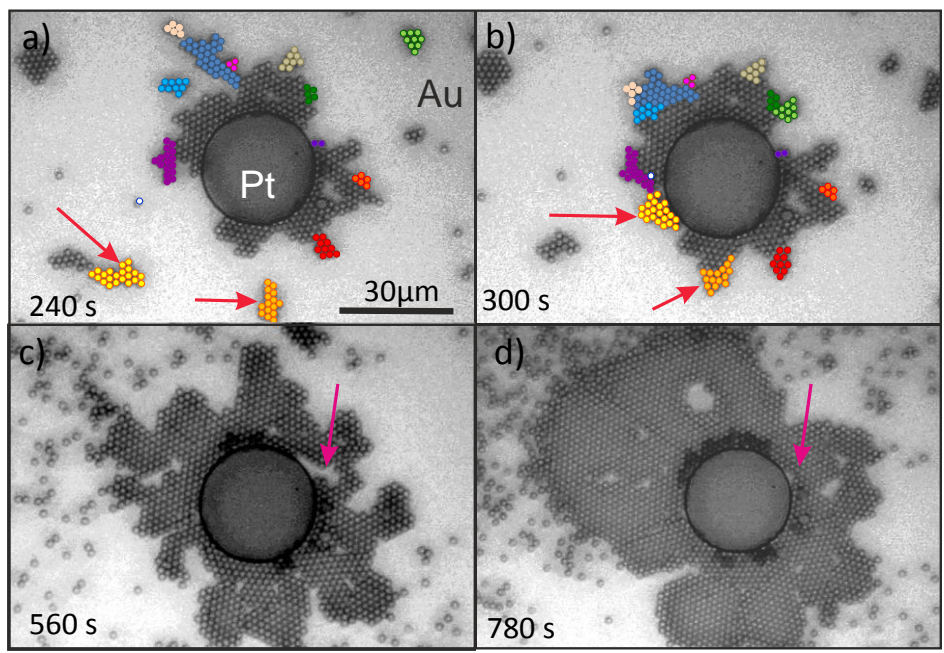

**Figure 5**. (a-d) Evolution of the colloidal crystal. In (a-b) some incoming clusters are colored to show that their shape can change once they reach the crystal. In (c-d) the arrows illustrate the defect healing effect. The darker zone just at the rim of the Pt disk in (d) corresponds to region consisting of two layers of colloids.

In these experiments, it is important to clean the pump surface with the oxygen plasma treatment. Without oxygen plasma treatment, we only observe the repulsion of the colloids from the pump disk. That is, the colloid clustering, the motion towards the pump disk, and the colloidal crystallization around the pump disk do not occur. This suggests that the oxygen plasma cleaning favors the production of protons when the pump is subjected to $H_2O_2$. Without oxygen plasma cleaning, the production of protons is weak and is not enough to change significantly the zeta potential of colloids.

We also carry out experiments with positively charged particles (polystyrene spheres functionalized with amidine groups). The particles move towards the Pt disk where they adsorb onto its surface [48]. The particles form a structure that is not well ordered. In other words, the particles do not form a colloidal crystal.

4. CONCLUSIONS

In this paper we demonstrate the use of bimetallic catalytic pumps made of noble materials to guide colloidal crystallization at precise locations without the need of an external energy source. The local self-generated electro-hydrodynamic forces induced by electrochemical reactions together with the sensitivity of the colloid zeta potential to the local pH conditions are the basic ingredients for tailoring the colloidal crystal process. Silica colloids crystalize in a two dimensional fashion through a series of sequential steps controlled by the proton concentration.

ASSOCIATED CONTENT

Supporting Information. Videos showing the different tasks performed by the catalytic micropumps. This material is available free of charge via the Internet at http://pubs.acs.org.

AUTHOR INFORMATION

**Corresponding Author**

*E-mail: MariaJose.Esplandiu@cin2.es.

ACKNOWLEDGMENT

We acknowledge support from the European Union (ERC-carbonNEMS project), the Spanish government (FIS2009-11284, MAT2012-31338), and the Catalan government (AGAUR, SGR).

TOC graphic

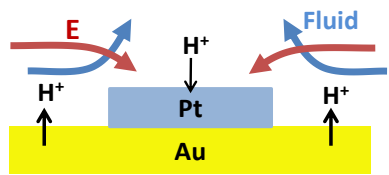

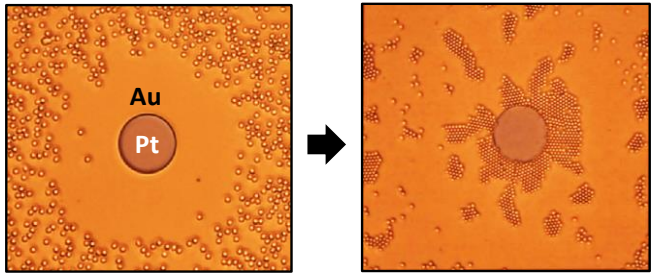